\documentclass[preprint,proceedings]{rmaa}

\newcommand{\starly}{{\sc Starlight}}
\newcommand{\niiha}{$\log\left([NII]\lambda 6584\mathrm{\AA}/H\alpha\right)$}
\newcommand{\oiiihb}{$\log\left([OIII]\lambda 5007\mathrm{\AA}/H\beta\right)$}
\newcommand{\ewnii}{$EW\left([NII]\lambda 6584\mathrm{\AA}\right)$}
\newcommand{\D}{\discretionary{}{}{}}

\SetYear{2010}
\SetConfTitle{XIII Latin American Regional IAU Meeting}

\title{Relation between activity, morphology and environment for a large sample of SDSS galaxies.}

\author{
R.~A.~Ortega-Minakata\altaffilmark{1},
J.~P.~Torres-Papaqui\altaffilmark{1},
H.~Andernach\altaffilmark{1},
J.~M.~Islas-Islas\altaffilmark{1}
and
D.~M.~Neri-Larios\altaffilmark{1}
}

\altaffiltext{1}{Departamento de Astronom\'ia, Universidad de Guanajuato;
Callej\'on de Jalisco S/N, Valenciana, C.P. 36240, Guanajuato, Gto., M\'exico
(rene\D{}@astro.\D{}ugto.\D{}mx).}

\suppressfulladdresses

\shortauthor{Ortega-Minakata et al.}
\shorttitle{Activity, Morphology and Environment of SDSS galaxies}

\listofauthors{}
\indexauthor{Ortega-Minakata,~R.~A.}
\indexauthor{Torres-Papaqui,~J.~P.}
\indexauthor{Andernach,~H.}
\indexauthor{Islas-Islas,~J.~M.}
\indexauthor{Neri-Larios,~D.~M.}

\addkeyword{Galaxies: active}
\addkeyword{Galaxies: statistics}

\begin{document}
\maketitle 

\boldabstract{We apply a stellar population synthesis code
to the spectra of a large sample of SDSS galaxies
to classify these according to their activity
(using emission-line diagnostic diagrams),
environment (using catalogues of isolated and cluster galaxies),
and using parameters that correlate with their morphology.
}

From the SDSS-DR7 \citep{dr7},
a flux-limited sample of galaxies was selected 
following selection criteria used by \citet[][-- Cid05]{cid05}.
We apply a stellar population synthesis code known as \starly~(Cid05)
to the spectra of the resulting sample (175111 galaxies), 
obtaining the  the intensity of the emission lines of each galaxy
(from the residual spectra).

Using emission-line diagnostic diagrams
(\citealp[][-- BPT]{bpt}; \citealp[][-- Coz98]{coz98}), 
and the diagnostic lines of \citet{kew01} and \citet{kau03},
we classified the sample in
three classes: Star Forming galaxies (SFs), AGN-hosts (AGNs) and
Transition Objects (TOs), classifying only galaxies with $S/N \geq 3$
in the relevant lines of each diagram.
Using the standard BPT diagram -- \niiha~vs. \oiiihb, the fractions of
galaxies in each class are $\sim$ 66\% SFs, 10\% AGNs and 24\% TOs, out of
91254 galaxies classifiable with this diagram. Combining this
diagram with the ``NII Diagram'' (Coz98) -- \niiha~vs. \ewnii,
the fractions are $\sim$ 55\% SFs, 23\% AGNs and 22\% TOs, out of 127827
galaxies classifiable with this diagram.

\citet{fuk07} found a correlation between the morphology and the
K-corrected photometric colours ($u-g$, $g-r$, $r-i$, and $i-z$,
defined in the SDSS photometric system), and also between the
morphology and the concentration index
$R_{50}(r)/R_{90}(r)$ (defined as the ratio of the 50\% and  90\%
Petrosian radii). Using these relations, we classified all the galaxies
in our sample adopting a ``morphological index''~T, ranging from 0 to 6
(0 = E, 1 = S0, 2 = Sa, 3 = Sb, 4 = Sc, 5 = Sd/Sm, 6 = Irr;
\citealp[see][]{fuk07}).

We constructed a subsample of galaxies that live in a
cluster environment, using the compilation of redshifts of
ACO \citep{aco} cluster member galaxies maintained by one of us
\citep[see][]{and05}. The figure shows the distribution
of morphological index T of galaxies in this subsample with different
activity types according to the NII Diagram. As a comparison, we show
the same distribution for 421 narrow-emission-line isolated galaxies
from \citet{kar10}. A strong morphological segregation between different
activity types can be seen in both environments.

\begin{figure}[!t]
  \includegraphics[width=\columnwidth]{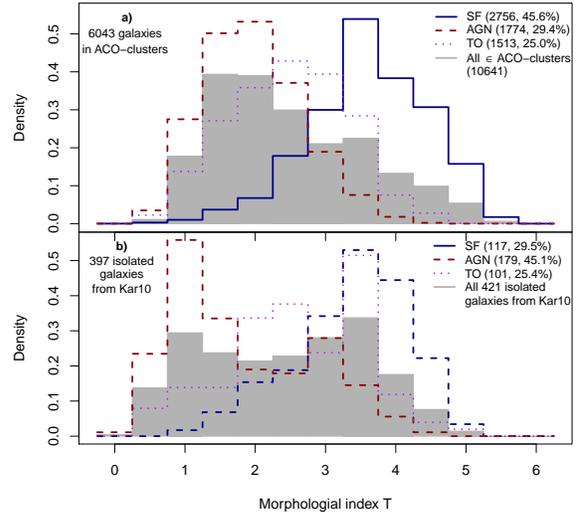}
  \caption{Morphological distribution of activity types for a) cluster galaxies,
    and b) isolated galaxies.\label{fig.morph_act}}
\end{figure}

\end{document}